\documentclass[fleqn,usenatbib]{mnras}

\usepackage{mathptmx}

\usepackage[T1]{fontenc}

\DeclareRobustCommand{\VAN}[3]{#2}
\let\VANthebibliography\thebibliography
\def\thebibliography{\DeclareRobustCommand{\VAN}[3]{##3}\VANthebibliography}

\usepackage{graphicx}	
\usepackage{amsmath}	
\usepackage{amssymb}	
\usepackage[version=3]{mhchem}
\usepackage{gensymb}
\usepackage{adjustbox}

\newcommand{\fig}{Fig.}
\newcommand{\Fig}{Fig.}
\newcommand{\figref}[1]{\fig~\ref{#1}}
\newcommand{\Figref}[1]{\Fig~\ref{#1}}

\newcommand{\Tabref}[1]{Table~\ref{#1}}

\renewcommand{\eqref}[1]{Eq.~(\ref{#1})}

\newcommand{\secref}[1]{Section~\ref{#1}}

\providecommand{\e}[1]{\ensuremath{\times 10^{#1}}}

\title[Nitrogen on ASW - II. Diffusion]{Neural-Network Assisted Study of Nitrogen Atom Dynamics on Amorphous Solid Water - II. Diffusion}

\author[Viktor Zaverkin et al]{
Viktor Zaverkin,$^{1}$
Germ\'an Molpeceres,$^{1}$
Johannes K\"astner$^{1}$\thanks{E-mail: kaestner@theochem.uni-stuttgart.de}
\\
$^{1}$Institute for Theoretical Chemistry,
University of Stuttgart,
Pfaffenwaldring 55, 70569 
Stuttgart, Germany\\
}

\date{Accepted XXX. Received YYY; in original form ZZZ}

\pubyear{2020}

\begin{document}
\label{firstpage}
\pagerange{\pageref{firstpage}--\pageref{lastpage}}
\maketitle

\begin{abstract}
The diffusion of atoms and radicals on interstellar dust grains is a fundamental ingredient for predicting accurate molecular abundances in astronomical environments. Quantitative values of diffusivity and diffusion barriers usually rely heavily on empirical rules. In this paper, we compute the diffusion coefficients of adsorbed nitrogen atoms by combining machine-learned interatomic potentials, metadynamics, and kinetic Monte Carlo simulations. With this approach, we obtain a diffusion coefficient of nitrogen atoms on the surface of amorphous solid water of merely $(3.5 \pm 1.1)\e{-34}$~cm$^2$s$^{-1}$ at 10 K for a bare ice surface. Thus, we find that nitrogen, as a paradigmatic case for light and weakly bound adsorbates, is unable to diffuse on bare amorphous solid water at 10~K. Surface coverage has a strong effect on the diffusion coefficient by modulating its value over 9--12 orders of magnitude at 10~K and enables diffusion for specific conditions. In addition, we have found that atom tunneling has a negligible effect. Average diffusion barriers of the potential energy surface (2.56~kJ~mol$^{-1}$) differ strongly from the effective diffusion barrier obtained from the diffusion coefficient for a bare surface (6.06~kJ~mol$^{-1}$) and are, thus, inappropriate for diffusion modeling. Our findings suggest that the thermal diffusion of N on water ice is a process that is highly dependent on the physical conditions of the ice.
\end{abstract}




\section{\label{sec:intro} Introduction}

The mobility of atoms and molecules on the surface of interstellar dust grains is crucial for surface processes like the formation of complex organic molecules. Observed abundances can be explained only by a combination of gas-phase reactions and surface chemistry~\citep{Herbst2009}. It is believed that diffusive processes, like the Langmuir--Hinshelwood mechanism, prevail in surface chemistry at low temperatures~\citep{Herbst2009,Ruaud2015}, although the Eley--Rideal and ``hot-atom'' mechanisms may have significant importance~\citep{He2017-b}.

The rate-limiting step of diffusive mechanisms is the mobility of adsorbates on the surface. At the average temperature of molecular clouds of 10--20~K~\citep{Snow2006}, diffusion by thermal hopping is limited.  The mass and binding energy are the main factors determining if an adsorbate diffuses or not. There is evidence from experiment~\citep{Tsong2001, Hama2012,Kuwahata2015} and simulation~\citep{sen17,asg17} for efficient diffusion of H, \ce{D}, \ce{H2} and \ce{He} on ice surfaces at temperatures as low as 10~K. Especially the diffusion of H may be facilitated by tunneling~\citep{Kuwahata2015} on polycrystalline water ice. On amorphous solid water (ASW), the influence of tunneling greatly depends on the adsorption site under consideration~\citep{Hama2012,sen17,asg17}.

A different situation arises for the next set of light particles with relevance in astrochemistry, namely the first-row atoms C, N, and O, for which rich chemistry is expected. The interaction of these atoms with ASW has been theoretically studied recently~\citep{Shimonishi2018}, finding that C forms a tightly bound complex and is unable to diffuse, while O and especially N are much weaker bound. Quantum tunneling was claimed by indirect evidence to be responsible for O diffusion~\citep{Minissale2013}, a finding disputed later~\citep{Pezzella2018}. Here, we report on the diffusivity of nitrogen atoms on amorphous solid water surfaces.

Recently, we investigated the adsorption dynamics of the N atom on ASW~\citep{Molpeceres2020_2} using ab-initio molecular dynamics employing a neural-network potential (MLP)~\citep{Zaverkin2020}, finding sticking to be extremely effective at low temperatures and desorption to occur in a window in between 23--28~K, in agreement with previous experiments~\citep{Minissale2016}. The average binding energy of N on ASW was found to be very small ($\sim$ 2.9~kJ~mol$^{-1}$, including zero-point vibrational energies), also in agreement with the average value provided in recent simulations~\citep{Shimonishi2018}. Our reported distribution of binding energies is also in accordance with the experimental values of the literature \citep{Minissale2016}, with a lower average value (by a factor of two) but a significant amount of binding sites in their provided range ($\sim$ 5.8~kJ~mol$^{-1}$), see Figure 2 of \cite{Molpeceres2020_2}. 

The small binding energy, in combination with the small mass of N and the possibility of simulating long time scales thanks to the MLP, have motivated us to explicitly study the diffusion of an adsorbate other than \ce{H} and \ce{D} with relevance to interstellar surface chemistry. Using a combination of accelerated sampling techniques~\citep{Laio2002,Parrinello08} to construct a 2D free-energy surface experienced by the N atoms and kinetic Monte Carlo simulations, we have estimated the diffusion coefficient of N as a function of the temperature, as well as other diffusion-related properties.

The paper has the following structure: first, we briefly introduce the relevant computational details, \secref{sec:methods}, and describe the obtained results, \secref{sec:resutls}. The justification of our finding is presented in \secref{sec:discussion} while a rationalization of previous results in the light of our recent simulations is discussed in the last section, \secref{sec:conclusion}. The results presented here have important implications for the chemistry of dense, interstellar clouds.

\section{Computational Details} \label{sec:methods}

The study of diffusion requires long time scales, short time steps in direct molecular dynamics, and a very accurate potential energy surface. We have achieved this by combining ab-initio molecular dynamics with a neural network potential (MLP) \citep{Zaverkin2020, Molpeceres2020_2}, free-energy sampling using metadynamics, and kinetic Monte Carlo (kMC) based on the minima and saddle points on the free-energy surface. 

\subsection{Machine-Learned Potential}

To create an accurate interatomic potential, we employed the Gaussian moment neural network approach (GM-NN) \citep{Zaverkin2020} recently developed in our group. Our MLP was fitted to a training set of energies and gradients at the PBEh-3c/def2-mSVP \citep{Grimme2015} level of $28,715$ structures with $3$ to $378$ atoms each. For more details on the data set employed in this work see elsewhere \citep{Molpeceres2020_2}. The training set structures are accessible free of charge \citep{MolpeceresDataset}. For details on the construction of the respective machine-learned potential (MLP) and on the analysis of its accuracy during molecular dynamics simulations, see Supplementary Information.

\subsection{Free Energy Sampling}

\begin{figure*}
    \centering
    \includegraphics[width=16cm]{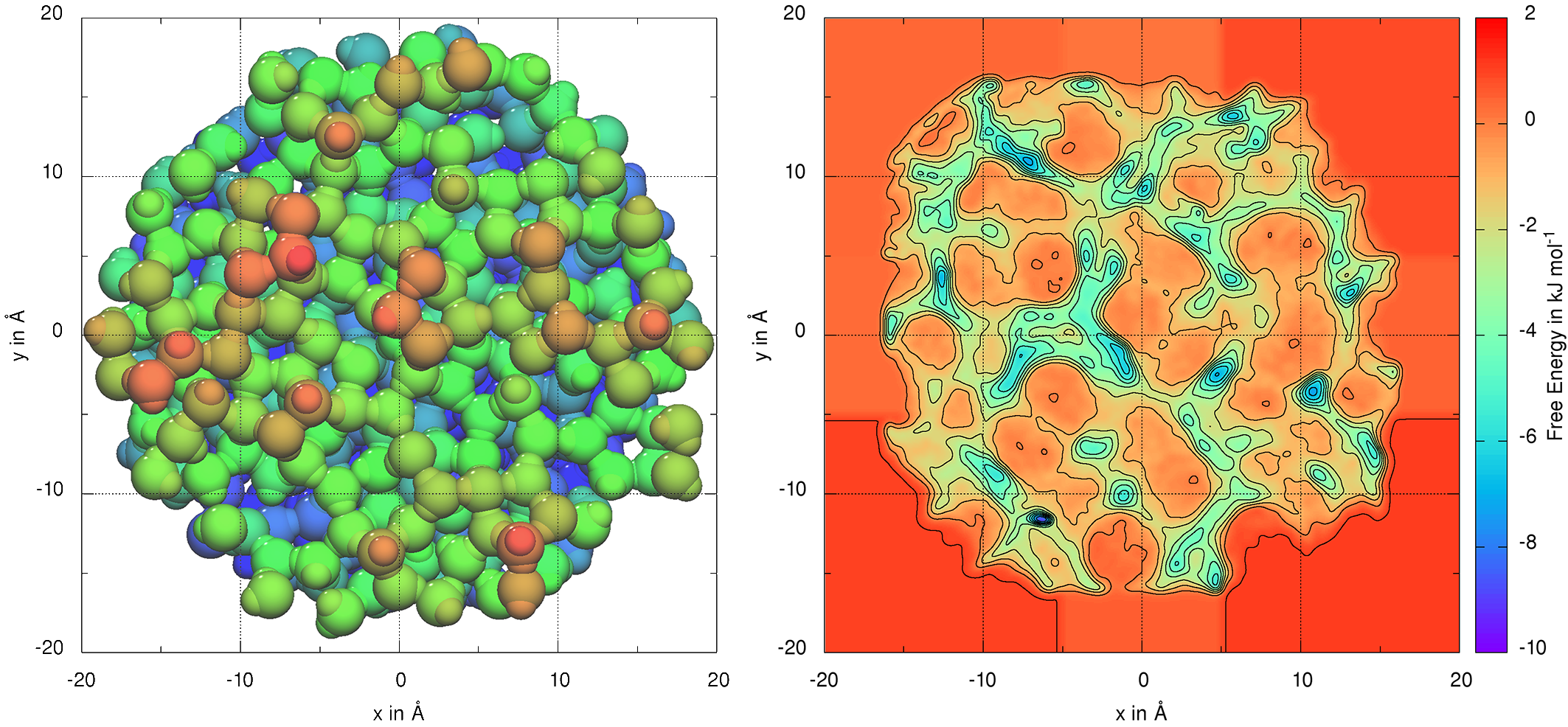}
    \caption{(Left) Atomic structure of the amorphous water ice equilibrated at 50~K. 
    All atoms are colored according to their $z$-coordinate value (surface normal) with red atoms sticking out of the surface and blue atoms denoting cavities. (Right) The 2D free-energy surface for the adsorbed nitrogen atom on the ice surface.}
    \label{fig:1}
\end{figure*}

The free-energy sampling of a region spanning 800~{\AA}$^2$ was performed using well-tempered metadynamics simulations \citep{hub94,Laio2002,Parrinello08}. For that, we interfaced our neural-network potential with the Atomic Simulation Environment (ASE) \citep{HjorthLarsen2017} and the PLUMED package \citep{Massimiliano09,Tribello14,plumed19}. The internal modes of water were flexible. The collective variables for the metadynamics were selected to be the $x$ and $y$ components of the nitrogen atom diffusing on the surface. The parameters of the Gaussian bias potential in the metadynamics were a rate of deposition of 125~fs, a Gaussian height of 0.025 kJ/mol, a Gaussian width of 0.25~{\AA} for each of the collective variables, and a bias factor of 6.
These parameters were obtained after extensive testing to produce a smooth free energy surface. In addition, we included an arbitrarily high wall potential (spherical) to avoid nitrogen escaping via the borders of the ice. The complete sampling of such a big region is impossible, even with metadynamics simulations. Thus, we have divided the complete surface into 9 different sub-regions of reduced size running metadynamics simulations in each one of them. Jumps between the different sub-regions were avoided by a harmonic potential wall. The FES was then reconstructed by overlapping each sub-region using a weighting function to ensure a smooth potential. For more details, see Supplementary Information. Each sub-region was sampled at a temperature of 50~K for a total time of 6.5~ns each. The free energy surface was reconstructed from the negative of the history-dependent bias potential \citep{Laio2002}. Our structural model of the ice and the reconstructed 2D free-energy surface for the movement of the nitrogen atom in $x$ and $y$ directions are shown in \figref{fig:1}.

\subsection{Minima and Transition State Optimization} \label{sec:coarse_grain}

Using the analytic free energy surface and its derivatives, we optimized the minima and transition states using the optimization library DL-FIND \citep{DLFIND09}. We obtained 139 minima and coarse-grained them to 60 by combining close-lying minima separated by negligible barriers. We calculated the transition states that can be reduced to a single elementary step using the NEB method \citep{NEB, Henkelman00_1, Henkelman00_2}. We ended up with 60 minima interconnected by 107 transition states. All relevant data are given in Supplementary Information, to ensure the reproducibility of this work. Rate constants for each possible transition (direct and reverse) were obtained in the context of transition state theory by applying Eyring's equation. Explicit consideration of tunneling for each transition was incorporated using Eckart and Bell-type tunneling corrections for each rate constant.

\subsection{Modeling of Diffusion}

We constructed a kinetic Monte-Carlo (kMC) model~\citep{Bortz75,Gillespie76} explicitly considering all the activation barriers and minima on our FES. We ran $10^8$ kMC steps at each temperature ($10^9$ steps at 17~K). After that, the probability of finding the N adatom at each binding site in kMC resembled its Boltzmann probability. This was achievable down to 17~K. To reach lower temperatures, we had to extrapolate, see \figref{fig:3}. To take our model's boundaries into account appropriately, we divided the diffusion paths into segments \citep{Kirchheim87, Kirchheim88, Ramasubramaniam08} that end once the N atom reaches the confining potential wall. A new segment is started if a random number between 0 and 1 is larger than $0.5$ to mimic the possible hop out of our boundaries. Additionally, we started a new kMC segment if, during $10^5$ iterations, no border was reached to obtain better statistics on estimated diffusion coefficients. The diffusion coefficient was calculated from the mean square displacement of the nitrogen atom by time-averaging over the segments~$i$
\begin{equation}
  D = \sum_i \frac{D_i \triangle t _i}{t},
\end{equation}
where $\triangle t_i = t_i - t_{i-1}$ is the time length of the segment $i$, $t$ is the total time of the kMC simulation, and the respective diffusion coefficient $D_i$ is calculated as
\begin{equation}
    D_i = \frac{\left(\mathbf{r}\left(t_i\right) - \mathbf{r}\left(t_{i-1}\right)\right)^T \left(\mathbf{r}\left(t_i\right) - \mathbf{r}\left(t_{i-1}\right)\right)}{2 d \triangle t_i},
\end{equation}
where $d=2$ for a two-dimensional system. We carefully validated all simulation parameters, also the non-periodic model, by comparison between easy-to-sample periodic and non-periodic auxiliary models, see Supplementary Information. Finally, the simulation of different degrees of surface coverage was done by removing specific minima for the kMC simulation, assuming a non-reactive species is already occupying such state, therefore the distribution of binding sites do not change by the number of adsorbed atoms.

The metadynamics simulations required direct molecular dynamics of  6.5~ns, while the kMC runs covered $10^{12}$~s. This protocol resulted in temperature-dependent diffusion coefficients as time averages of our kMC trajectories.

\section{Computational Results} \label{sec:resutls}

\subsection{\label{sec:fes} Free Energy Surface}

\begin{figure*}
    \centering
    \includegraphics[width=16cm]{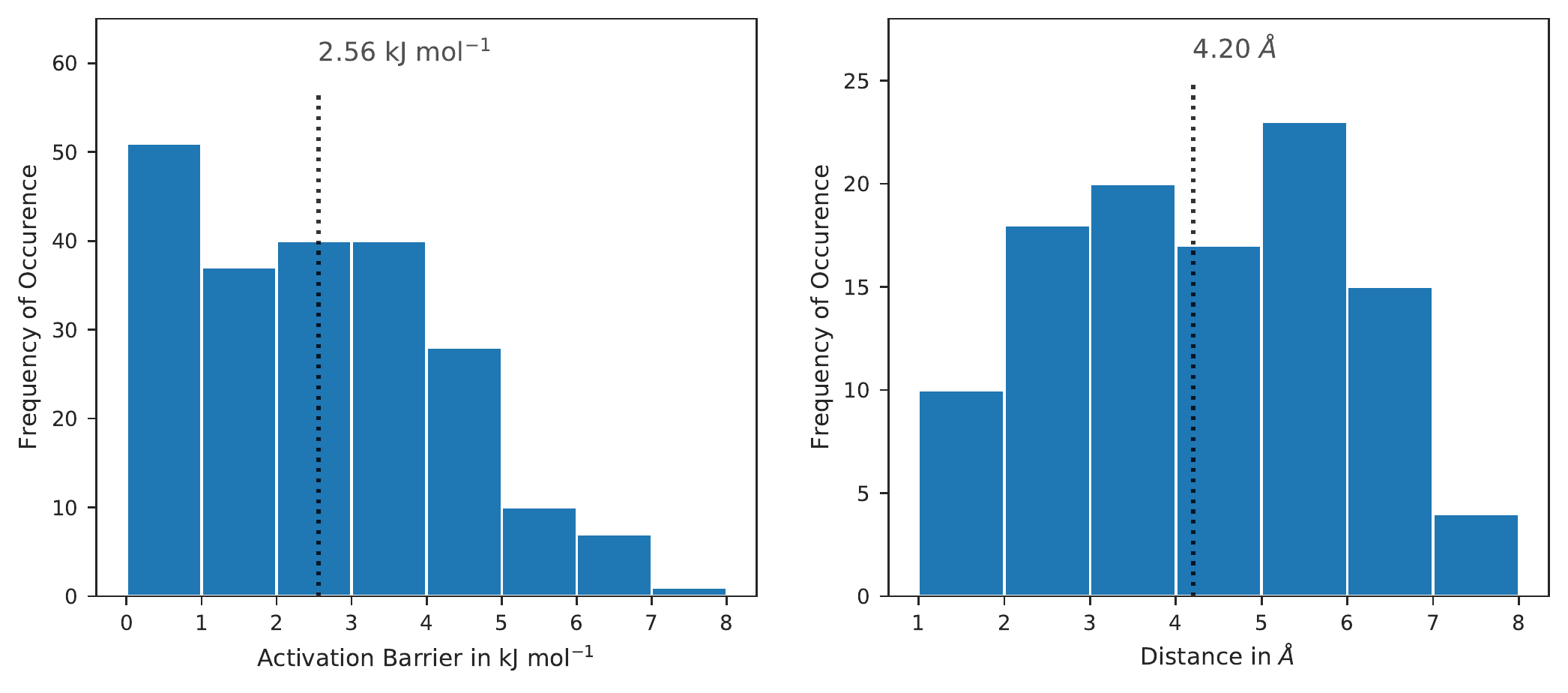}
    \caption{(Left) Distribution of diffusion barriers with a mean of $2.56$~kJ~mol$^{-1}$ and a standard deviation of $1.72$~kJ~mol$^{-1}$. (Right) Distribution of distances between neighboring sites on the FES with a mean of $4.20$~\AA{} and a standard deviation of $1.63$~\AA{}.}
    \label{fig:2}
\end{figure*}

\figref{fig:1}~(left) shows the amorphous solid water (ASW) ice surface equilibrated at 50~K with atoms colored according to their $z$-coordinate. \figref{fig:1}~(right) represents the respective 2D free energy surface (FES) for the adsorbed nitrogen atom on the ice surface. From these figures, the complexity of the ASW and FES surface topologies can be deduced. Moreover, the presence of small pores is expected to be crucial when computing the diffusion coefficients. Note that we assume that differences between FESs obtained for temperatures equal to or lower than 50~K are small since the variation of the entropy of the system should be minute. Therefore, it is sufficient to sample the FES at $50$~K only and use the obtained diffusion barriers within the kinetic Monte Carlo framework for other temperatures.


\Figref{fig:2}~(left) shows the distribution of activation barriers for diffusion obtained for the coarse-grained free-energy surface. It can be seen that the respective distribution is relatively broad with a mean value of $2.56$~kJ~mol$^{-1}$ and a standard deviation value of $1.72$~kJ~mol$^{-1}$. Taking into account the broad distribution of activation energies, one could claim that the assumption about the relation of the effective diffusion barrier to the mean binding energy, i.e., $E_\mathrm{diff} \sim 0.55 E_\mathrm{bin}$ found in the literature~\citep{Minissale2016}, depends on the physical conditions under consideration since it does not directly compare with our value of 0.76 (computed from our distribution of binding energies excluding zero-point vibrational energies, see \cite{Molpeceres2020_2}). 

In addition, one may estimate the pre-exponential factor, $D_0$, of the classical Arrhenius expression
\begin{equation}
\label{eq:arrhenius}
    D = D_0 \exp\left(-\frac{\Delta F}{RT}\right),
\end{equation}
directly from FES. For this purpose one may write for $D_0$, similar to \cite{Du2012},
\begin{equation}
\label{eq:d0}
    D_0 = \Gamma a_0^2 \nu_0,
\end{equation}
to which we add a superscript, i.e. we write $D_0^\mathrm{avg}$, to simplify the comparison of the obtained result to the corresponding values presented in \secref{sec:kmc}. In the expression in \eqref{eq:arrhenius}, we set the effective diffusion barrier $\Delta F$ to $\Delta F^\mathrm{avg} = 2.56$~kJ~mol$^{-1}$ and, in \eqref{eq:d0}, $a_0$ is the mean jump distance, $\nu_0$ is the attempt frequency, and $\Gamma$ is the geometric pre-factor related to the connectivity of each site to its neighboring sites.

The attempt frequency $\nu_0$ can be estimated as the vibrational frequency of the adatom averaged over all sites. We obtained a value of $\nu_0 = 8.9\cdot 10^{12}$~s$^{-1}$. The respective harmonic frequencies are calculated from the Hessian matrix computed for the binding sites on the coarse-grained FES. 

The jump distance can be estimated from the distance distribution between neighboring minima, shown in \figref{fig:2} (right). Thus, we obtain a mean jump distance value of $a_0 = 4.20$~{\AA}. For the estimation of geometric pre-factor $\Gamma$ we assume isotropic diffusion, which results in the following expression \citep{Allnatt93}
\begin{equation}
    \Gamma = \frac{n}{2 d},
\end{equation}
where $n$ is the number of neighboring states and $d$ is the dimensionality of the system. The former can be determined as the mean of the connectivity of all sites and equals to $4$. Thus, we obtained the pre-exponential factor of $D_0^\mathrm{avg} = 1.57 \e{-2}$~cm$^2$s$^{-1}$, setting $d=2$. The respective Arrhenius plot is depicted in \figref{fig:3}.

Note that due to the broad distribution of activation barriers, distances between neighboring minima, and the number of neighboring states the quantities derived in this section, $D_0^\mathrm{avg}$ and $\Delta F^\mathrm{avg}$, may deviate from the ground truth. Thus, a more rigorous description of diffusion processes is needed. The Kinetic Monte Carlo approach, employed in \secref{sec:kmc}, provides us with the necessary flexibility taking into account the connectivity between binding sites with their realistic barriers.


\subsection{Kinetic Monte Carlo} \label{sec:kmc}

In \secref{sec:fes} we have seen that the 2D free-energy surface for the adsorbed nitrogen atom has a broad distribution of diffusion barriers, distances between neighboring sites, and the number of neighbors. Therefore, to avoid hopping between two states separated by a low activation barrier and, thus, improve the efficiency of kinetic Monte Carlo (kMC) simulations, we employ the coarse-grained FES, the generation of which is briefly discussed in \secref{sec:coarse_grain}.

\subsubsection{Single Adsorbed Nitrogen} \label{sec:bare surface}

To study the mobility of the nitrogen atom on the water surface, we performed kMC simulations at six different substrate temperatures $T = \{17, 20, 25, 30, 40, 50\}$~K. Each simulation was started from a randomly selected site and was performed for $10^8-10^9$ steps resulting in total times ranging from seconds ($50$~K) to several thousands of years ($17$~K). For each temperature value, we performed $25$ independent kMC runs with the exception of $T=17$~K for with 10 independent kMC runs were performed. The reason for that is the increased computational cost compared to other temperature values due to the increased number of steps ($10^9$). Converging kMC simulations at $10$~K was impossible in a reasonable time. 

To make sure that the sampling was performed long enough, we compared kMC probability to find the nitrogen atom at a certain binding site to its Boltzmann equivalent. The kMC probability can be calculated as $p_i \propto t_i/t_\mathrm{total}$, where $t_i$ is the time spend in the corresponding binding site $i$ and $t_\mathrm{total}$ is the total time covered by the simulation. 

\begin{figure}
        \centering
        \includegraphics[width=8cm]{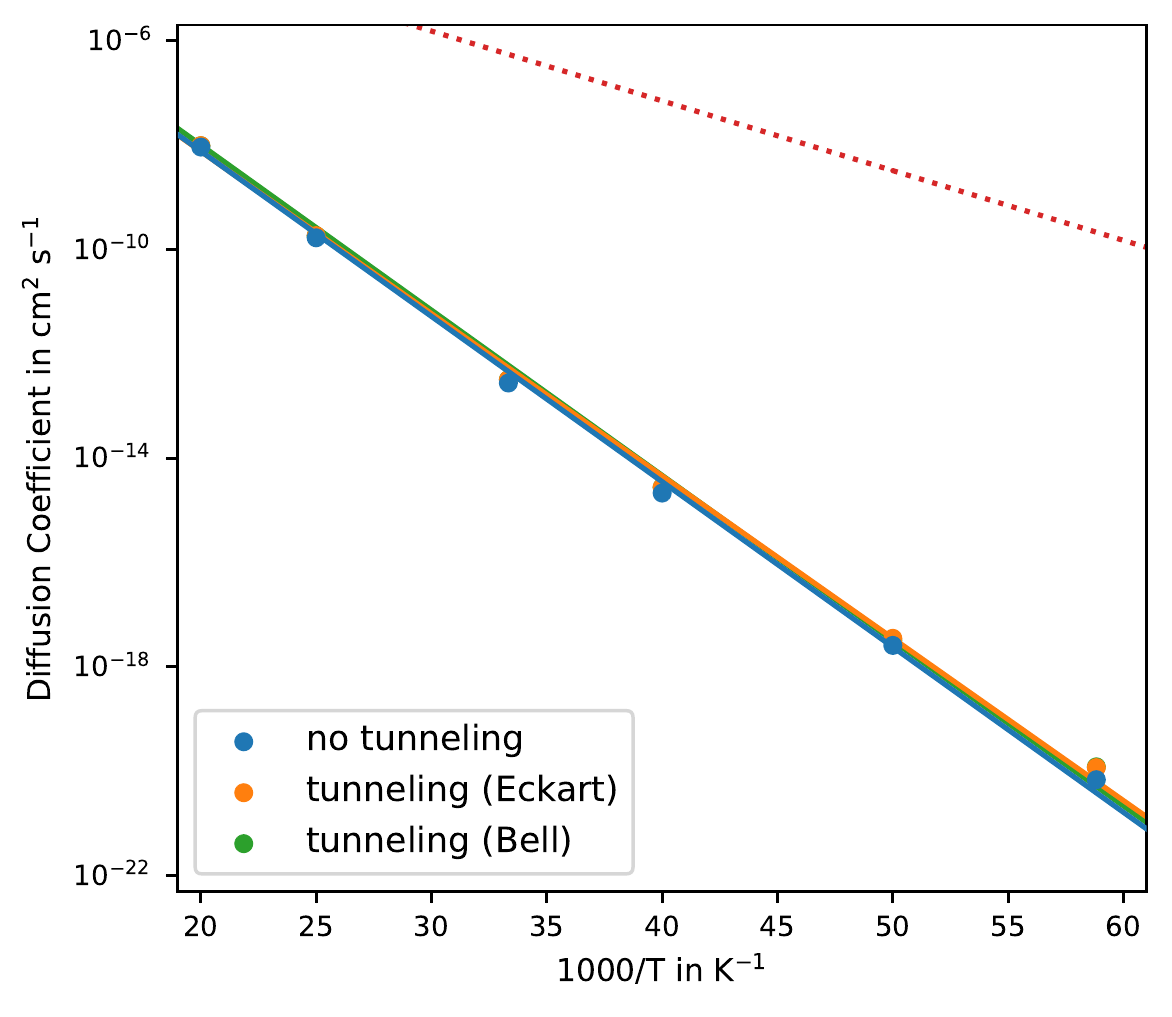}
    \caption{Temperature-dependence of diffusion coefficients ($D$) for N on ASW for the bare surface with or without tunneling correction. Linear fits are displayed for clarity, and the red dotted line represents $D^\text{avg}$ estimated in \secref{sec:fes}.}
    \label{fig:3}
\end{figure}

\figref{fig:3} shows the temperature-dependent diffusion coefficients obtained for the bare surface employing kMC simulations along with the respective linear fit. Fitting the respective values by the Arrhenius expression presented in \eqref{eq:arrhenius} we obtained the pre-exponential factor value of $D_0 = (1.65 \pm 0.32)\e{-2}$~cm$^2$s$^{-1}$ and the effective diffusion barrier value of $\Delta F = (6.06 \pm 0.04)$~kJ~mol$^{-1}$. It should be noted that diffusion coefficients predicted employing kMC simulations are different from $D^\text{avg}$, obtained using the averaged diffusion barrier in \secref{sec:fes}, by several orders of magnitude. While the pre-exponential factor is close to the one obtained in \secref{sec:fes} ($D_0^\mathrm{avg} = 1.57 \e{-2}$~cm$^2$s$^{-1}$), the average diffusion barrier ($\Delta F^\mathrm{avg} = 2.56$~kJ~mol$^{-1}$) is about $2.4$ times smaller compared to the respective value obtained by kMC simulations. This makes the diffusion at low temperatures less probable compared to the estimations based on the FES only. 

Using the data from the linear fit in \figref{fig:3} we estimated the diffusion coefficient at $10$~K and obtained $D\left(10~\mathrm{K}\right)=(3.47 \pm 1.07)\e{-34}$~cm$^2$s$^{-1}$. This is much lower than the estimate obtained using parameters from \secref{sec:fes} ($D^\mathrm{avg}\left(10~\mathrm{K}\right)=6.67\e{-16}$~cm$^2$s$^{-1}$). All numerical values of diffusion coefficients are presented in \Tabref{tab:1}.

\begin{table*}
	\centering
	\caption{Diffusion coefficients in cm$^2$s$^{-1}$ obtained for the bare surface, including tunneling and with the 1--4 deepest sites blocked by running kMC simulations. Standard deviation is given in parentheses. Values for $10$~K are estimated from the linear fit.}
	\label{tab:1}
	\begin{adjustbox}{max width=\textwidth}
    \begin{tabular}{cccccccc}
    \hline
    $T$ (K) & bare surface & tunneling (Eckart) & tunneling (Bell) & 1 site blocked & 2 sites blocked & 3 sites blocked & 4 sites blocked\\
    \hline
    10 & 3.47\e{-34} (1.07) & 9.82\e{-34} (2.79) & 4.67\e{-34} (1.98) & 8.20\e{-26} (3.30) & 2.87\e{-25} (0.33) & 9.71\e{-24} (2.26)& 9.04\e{-23} (1.86)    \\
    17 & 6.87\e{-21} (2.22) & 1.17\e{-20} (0.26) & 1.21\e{-20} (0.33) & 5.06\e{-16} (1.93) & 5.47\e{-16} (0.76)& 4.56\e{-15} (1.30)& 1.47\e{-14} (0.35)     \\
    20 & 2.57\e{-18} (0.58) & 3.52\e{-18} (0.98) & 3.22\e{-18} (1.55) & 1.96\e{-14} (0.67) & 2.98\e{-14} (0.12)&  1.59\e{-13} (0.13)&  4.69\e{-13} (0.49)   \\
    25 & 2.14\e{-15} (0.11) & 2.75\e{-15} (0.19) & 2.77\e{-15} (0.18) & 1.97\e{-12} (0.15) & 3.61\e{-12} (0.04) &  1.27\e{-11} (0.02) &  2.96\e{-11} (0.07) \\
    30 & 2.75\e{-13} (0.11) & 3.20\e{-13} (0.09) & 3.20\e{-13} (0.12) & 6.77\e{-11} (0.58) & 1.12\e{-10} (0.02) & 2.95\e{-10} (0.06) &  5.88\e{-10} (0.15)  \\
    40 & 1.67\e{-10} (0.02) & 1.84\e{-10} (0.02) & 1.82\e{-10} (0.02) & 7.17\e{-9} (0.13) & 1.11\e{-8} (0.03)& 2.08\e{-8} (0.06) & 3.32\e{-8} (0.09)        \\
    50 & 9.16\e{-9}  (0.04) & 9.74\e{-9} (0.05) & 9.71\e{-9} (0.06)   & 1.35\e{-7} (0.03) & 1.94\e{-7} (0.03) & 3.02\e{-7} (0.04) & 4.26\e{-7} (0.06)       \\
    \hline
    $D_0$      & 1.65\e{-2} (0.32)  & 1.27\e{-2} (0.23) & 2.16\e{-2} (0.57) & 4.88\e{-3} (1.23) & 3.69\e{-3} (0.27) & 3.02\e{-3} (0.44) & 2.63\e{-3} (0.34) \\
    $\Delta F$ & 6.06 (0.04)        & 5.96 (0.04)       & 6.06 (0.06)       & 4.36 (0.05)       & 4.23 (0.02)       & 3.92 (0.03)       & 3.73 (0.03)       \\
    \hline
    \end{tabular}
    \end{adjustbox}
\end{table*}

Note that it is advisable to perform companion molecular dynamics simulations to ensure that kMC simulations lead to correct state-to-state evolution of the studied system \citep{Voter2007}. However, many theoretical studies have been performed on similar systems~\citep{Karssemeijer2013, Karssemeijer2014, Senevirathne2017, asg17} like the molecular dynamics study of \cite{Pezzella2018} which we use as a reference in this work.

From Figure 2 A of \cite{Pezzella2018} one finds a mean squared displacement (MSD) of 133.64~\AA$^2$ for the adsorbed oxygen atom sampled over 500~ns at 50~K. This results in a diffusion coefficient of 6.68\e{-9}~cm$^2$ s$^{-1}$ similar to the one obtained by us for the nitrogen atom (9.16\e{-9}~cm$^2$ s$^{-1}$). Similar values for the diffusion coefficients might be expected because the estimated average diffusion barriers are close for both systems and are 2.29~kJ~mol$^{-1}$ \citep{Pezzella2018} for the oxygen atom and 2.56~kJ~mol$^{-1}$ for the nitrogen atom. 

Additionally, we analyzed direct molecular dynamics simulations of the nitrogen atom adsorbed on top of the ASW surface performed over 4~ns and 2~ns at 10~K and 50~K, respectively. The analysis revealed the good correspondence of the diffusion coefficients obtained by the direct molecular dynamics and by our kMC simulations. For details see Supplementary Information. Based on these results we may argue that the kMC dynamics produces correct time evolution of the system and, thus, is statistically indistinguishable from a long molecular dynamics simulation.

Quantum tunneling may potentially influence the mobility of nitrogen atoms. We use the Eckart and Bell corrections to the rate constants of the kMC simulations. For analytic expressions of the respective tunneling corrections see \cite{McConnell17}. 
The diffusion coefficients obtained using the corresponding corrected rate constants are shown in \figref{fig:3}. The corresponding numerical values, as well as $D_0$ and $\Delta F$, can be found in \Tabref{tab:1}.

From \figref{fig:3}, we see that the diffusion coefficients accounting for tunneling are only marginally larger compared to the ones without tunneling corrections. This result can be expected, taking into account the high mass of the nitrogen atom.

Another concept that has to be accounted for is the roughness of the underlying potential at the atomic length scales as introduced by \cite{Pezzella2018}. Following the work of \cite{Zwanzig1988} we may write for the effective diffusion coefficient 
\begin{equation}
    D^\ast = D \exp{\left(-\left(\frac{\epsilon}{RT}\right)^2\right)}, 
\end{equation}
where we use the Arrhenius expression for $D$ from \eqref{eq:arrhenius} and $\epsilon$ resembles the variations of the potential on atomic length scales.
Note that in contrast to \cite{Pezzella2018} we obtain a modified expression
\begin{equation}
    \label{eq:non_arrhenius}
    D^\ast = D_0 \exp{\left(-\frac{\Delta F}{RT} - \left(\frac{\epsilon}{RT}\right)^2\right)}, 
\end{equation}
which is more suitable for fitting properties which vary on several orders of magnitude like the diffusion coefficient.

Fitting the expression in \eqref{eq:non_arrhenius} to the kMC data results in a negligibly small roughness parameter $\epsilon$ and an increase of diffusion coefficient at 10~K of a factor of 2. Therefore, we skip the respective analysis in the subsequent sections.

\subsubsection{Blocking Surface Sites} \label{sec:site_blocking} 

In general, the studied water surface has regions containing smaller pores on different parts of the ice surface that are characterized by high binding energy and a high number of neighbors. Note that in real ASW, much deeper pores are expected \citep{Bossa2015}. While performing kMC simulations on the bare surface, see \secref{sec:bare surface}, we observed that the nitrogen atom is able to diffuse quickly into one of these sites. Once the nitrogen atom was trapped by one of these sites, it stays there for most of the time. We have found that the nitrogen atom spends about $93$~\% of the simulation time in the site associated with the deepest free energy at $50$~K, found at around $(x, y) = (-6.3, -11.6)$~\AA{} for the respective free energy surface shown in \figref{fig:1}. For lower temperatures, the nitrogen atom stays in the corresponding site up to $99.99$~\% of the total simulation time. The smallest barrier to get out of the respective minimum is $\Delta F = 5.90$~kJ~mol$^{-1}$.

Moreover, in an experimental setup or interstellar clouds, single nitrogen atom diffusion is highly unlikely due to the high atomic fluxes employed in the former and the relative molecular abundances, e.g. with respect to \ce{H2}, \citep{Ruaud2016} in the latter. Higher surface coverage is expected in both cases. This raises the question of the influence of the presence of additional inert species on the mobility of the adsorbed nitrogen atom. To address this question, we exclude the deepest binding sites (1--4), which mimics their occupation by some inert chemical species. Note that in this section, we neglect tunneling, since we have found that it has only a marginal impact on the mobility of the nitrogen atom on the water ice surface.

Similar to \secref{sec:bare surface} we performed kMC simulations at different substrate temperatures ranging from $17$~K to $50$~K. All kMC simulations were performed for $10^6$--$10^8$ steps since the Boltzmann distribution could be achieved faster when excluding the deeper sites. For all temperatures, $25$ independent kMC runs were performed. Simulations at $10$~K were impossible in a reasonable time due to computational cost even after excluding 4 deeper binding sites.

\begin{figure}
        \centering
        \includegraphics[width=8cm]{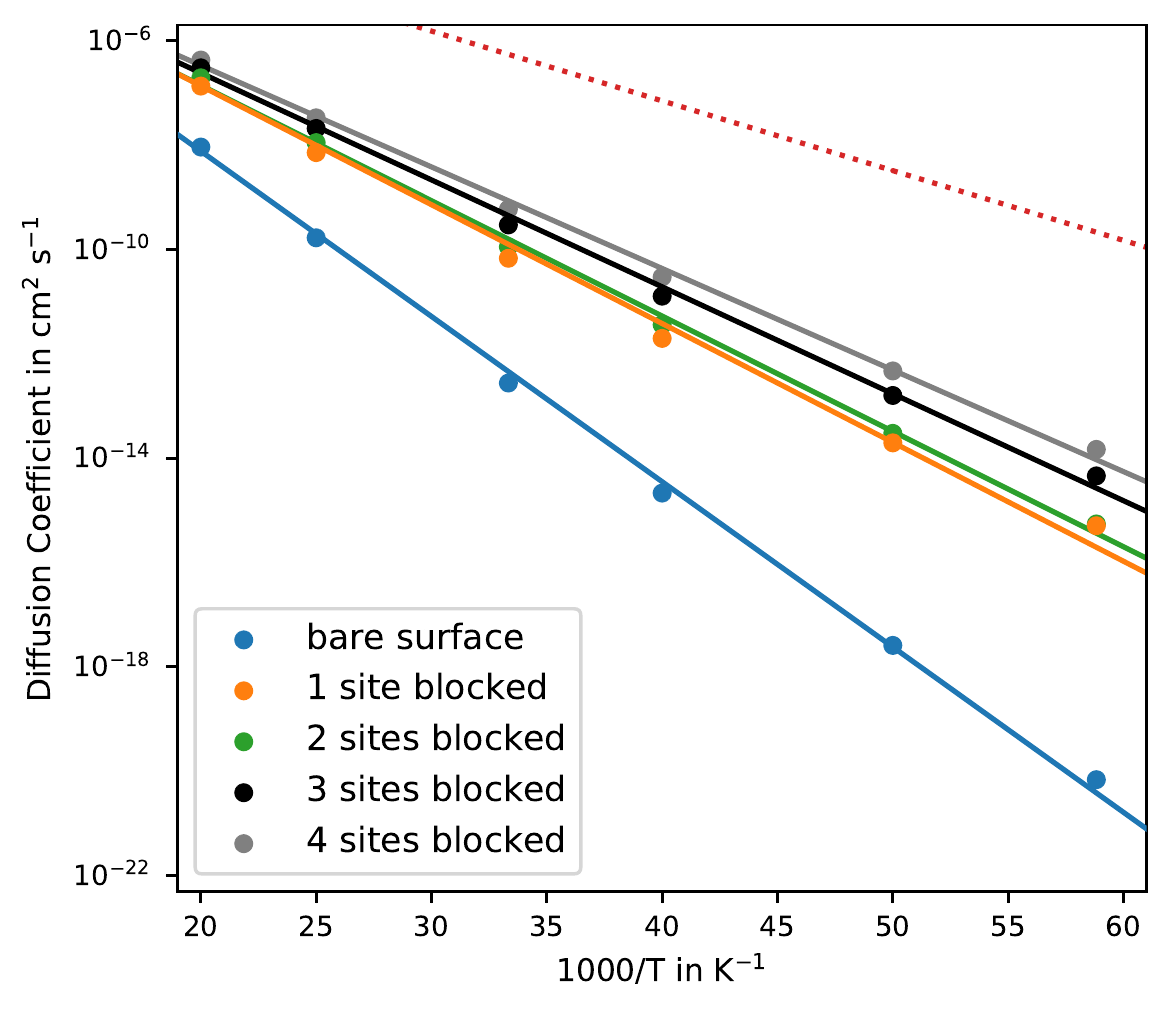}
    \caption{Temperature-dependence of diffusion coefficients ($D$) for N on ASW for the bare surface and the surface with the 1--4 deepest sites blocked. Linear fits are displayed for clarity, and the red dotted line represents $D^\text{avg}$.}
    \label{fig:4}
\end{figure}

\figref{fig:4} shows the temperature-dependent diffusion coefficients of the nitrogen atom along with the corresponding linear fits for the bare surface and the surface with 1--4 binding sites excluded from the simulation. From the figure, one can see that the mobility of the nitrogen atom on the water ice surface increases strongly with the increasing number of blocked sites. The strongest effect was observed by excluding the first deepest site. 
While the difference of diffusion coefficients for the bare surface and the surface with multiple occupied binding sites is smaller for temperatures above $30$~K, the effect on the mobility of the nitrogen is greater at lower temperatures. We have found that the diffusion coefficient for the adsorbed nitrogen atom at 10~K is 9 to 12 orders of magnitude larger for the surface with the blocked deeper binding sites compared to the bare surface. All numerical values of diffusion coefficients at different substrate temperatures can be found in \Tabref{tab:1}. 

\begin{figure}
        \centering
        \includegraphics[width=8cm]{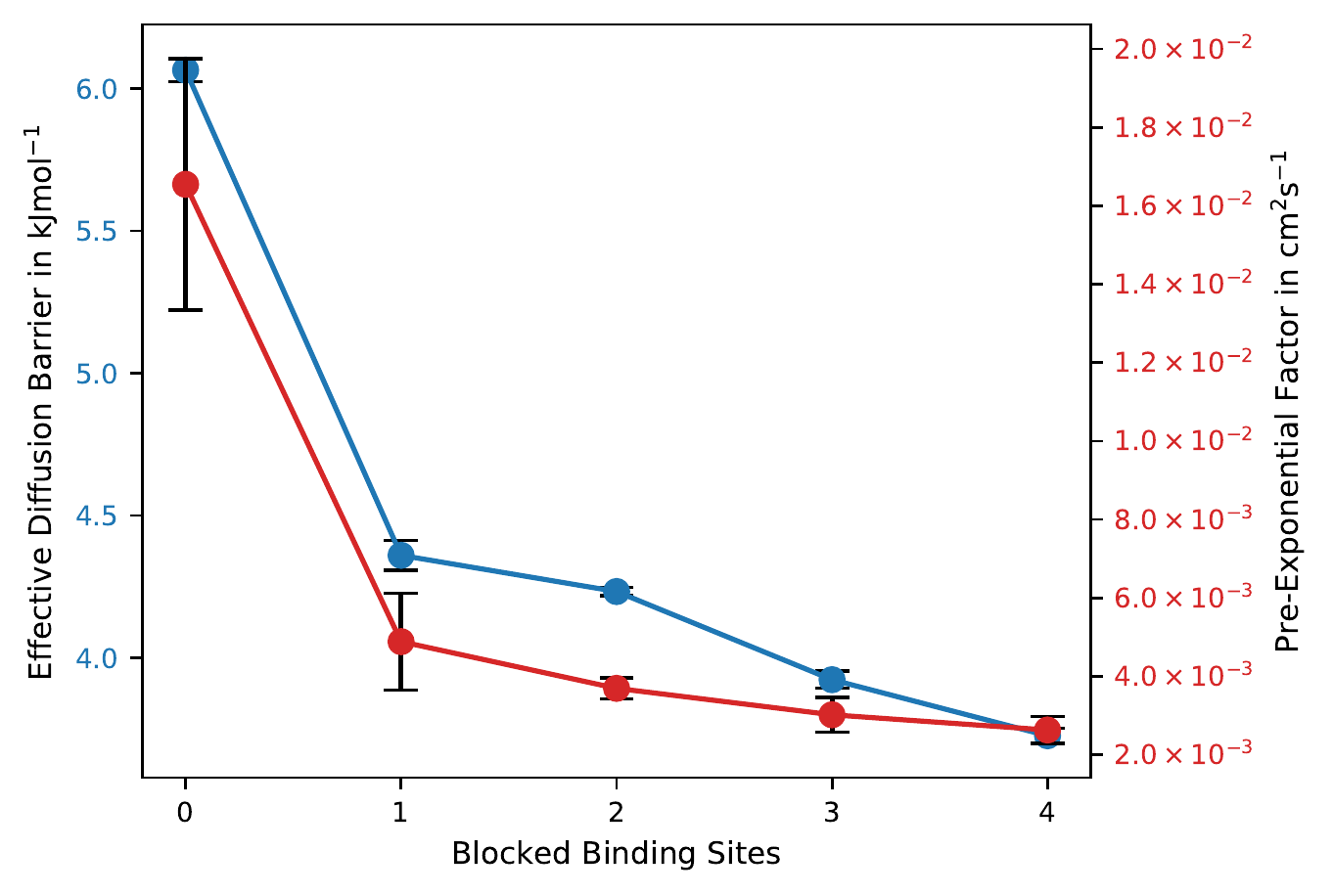}
    \caption{Pre-exponential factor, $D_0$, and effective diffusion barrier, $\Delta F$, depending on the number of blocked sites. The blue line shows the effective diffusion barrier, while the red one represents the pre-exponential factor.}
    \label{fig:5}
\end{figure}

\figref{fig:5} shows the dependence of the pre-exponential factor $D_0$ and the effective barrier $\Delta F$, obtained by fitting kMC values of diffusion coefficients by the Arrhenius equation from \eqref{eq:arrhenius}, on the number of occupied binding sites. The respective numerical values can be found in \Tabref{tab:1}.
From the figure, we see that the strongest decrease in both values was achieved already by excluding the deepest minimum in agreement with the observed increase of diffusion coefficients discussed above. The effective diffusion barrier decreases from $6.06$~kJ~mol$^{-1}$ to $4.36$~kJ~mol$^{-1}$, which is now closer to the averaged one obtained in \secref{sec:fes} ($\Delta F^\text{avg} = 2.56 \pm 1.72$~kJ~mol$^{-1}$)
, especially taking into account its large standard deviation. The pre-exponential factor decreases by one order of magnitude and remains almost unchanged after excluding the first binding site.

\subsubsection{Diffusion Times} \label{sec:diffusion_times} 

\begin{figure}
        \centering
        \includegraphics[width=\linewidth]{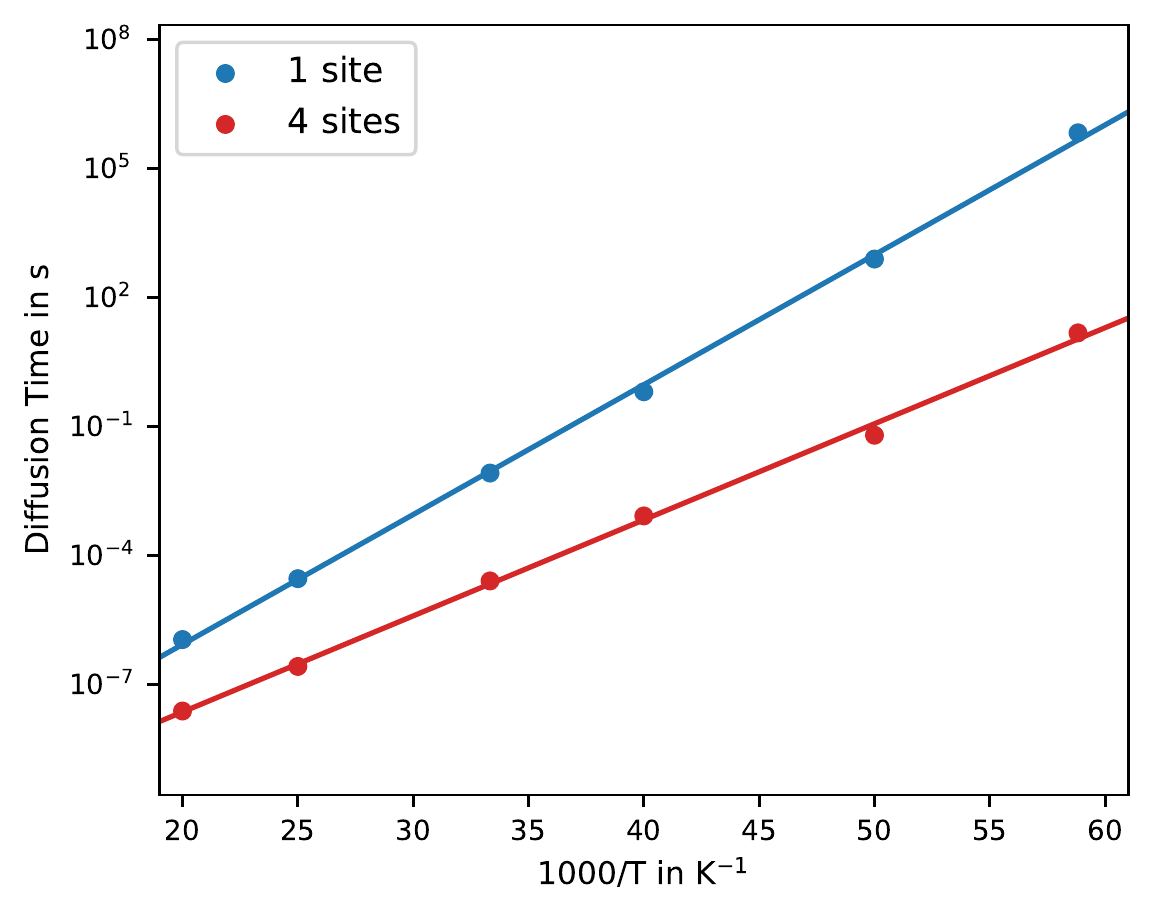}
    \caption{Diffusion times of the nitrogen atom adsorbed on top of the ASW surface until it is trapped by the deepest binding site (1 site) or by one of the four deepest binding sites (4 sites).}
    \label{fig:6}
\end{figure}

Besides diffusion coefficients, we measured times the nitrogen could freely diffuse on the bare surface before being trapped by the deepest binding site or by one of the four deepest binding sites. \figref{fig:6} shows the temperature dependence of the diffusion time for both scenarios. We have found that the nitrogen atom is trapped by the deepest binding site 
already after $1.10\e{-6}$~s at 50~K and after about $8$~days at $17$~K. The diffusion time depends exponentially on the inverse temperature, i.e., we may write
\begin{equation}
    t = A \exp\left(\frac{B}{T}\right)~,
\end{equation}
and obtain parameters $A$ and $B$ by fitting the respective expression to measured data. The corresponding linear fits are presented in \figref{fig:6}. For the nitrogen being trapped by the deepest binding site, we obtained $A=(7.61 \pm 3.05)\e{-13}$~s and $B=(696.01 \pm 9.95)$~K. For the nitrogen being trapped by one of four deepest binding sites, we obtained $A=(7.65 \pm 3.53)\e{-13}$~s and $B=(514.79 \pm 11.47)$~K.

Using the parameters $A$ and $B$, we can compute the time the nitrogen atom will be able to freely diffuse until it reaches the deepest or one of four deepest binding sites. The respective values are $1.28\e{18}$~s and $1.74\e{10}$~s (age of the Universe is $4\e{17}$~s). This is equivalent to effective distances passed by the nitrogen atom ($4Dt = r^2$) of 4.22~\AA{} and $5\e{-4}$~\AA{}, respectively. While the times are large, these distances are negligibly small compared to the size of typical dust grain of 1~$\mu$m.



\section{Discussion} \label{sec:discussion}

Running the kMC simulations at $T=50$ K, we obtain a diffusion coefficient of a nitrogen atom on ASW of $9.2\e{-9}$~cm$^2$~s$^{-1}$, which is already low, but still even higher than the experimental results \citep{Mate2020} obtained for \ce{CH4}, a molecule with very similar weight and affinity with the ASW to N, of 10$^{-12}$~cm$^2$~s$^{-1}$. The influence of the temperature is severe, at $T=17$ K we obtain $D=(6.9\pm 2.2)\e{-21}$~cm$^2$~s$^{-1}$, which is in much better agreement with estimates by He et al. \citep{He2018} of 10$^{-20}$~cm$^2$~s$^{-1}$ for \ce{CH4} on ASW. 
Extrapolating to $T=10$ K, the typical temperature in a molecular cloud, we arrive at $D=(3.5 \pm 1.1)\e{-34}$~cm$^2$~s$^{-1}$, a very low value. Since the mean-square displacement covered by diffusion in two dimensions is $\langle r^2\rangle=4Dt$, the average time for such a particle to scan a 1 $\mu$m dust grain is approx $10^{25}$~s, which is orders of magnitude longer than the age of the Universe ($4\e{17}$ s). Thus, a Langmuir--Hinshelwood mechanism of nitrogen atoms diffusing to meet reaction partners is found to be unlikely. 


The temperature-dependence of the diffusion constant nicely follows an Arrhenius-like behavior of $D(T)=D_0\exp(-\Delta F/RT)$ with $D_0=\left(1.65 \pm 0.32\right)\e{-2}$~cm$^2$~s$^{-1}$ and $\Delta F=6.06 \pm 0.04$~kJ~mol$^{-1}$, see \figref{fig:3}.

A closer analysis of our data reveals that the low diffusivity is caused by the domination of the strongest binding sites, as has been extensively discussed in \cite{Karssemeijer2014, sen17, asg17}. Whenever the N adsorbate finds one of the deep binding sites, it stays there for a long time. Our surface model is, with about 800~\AA$^2$, comparably small and certainly very compact and smooth. In real ASW, much deeper pores are expected, which should lead to even stronger binding. Thus, on a more realistic surface, the diffusion can be expected to be even slower. However, deep sites may be blocked by other adsorbates. That can be easily simulated with our model. Blocking 1 to 4 of the deepest sites results in much higher values for $D$ of $8.2\e{-26}$ to $9.0\e{-23}$~cm$^2$~s$^{-1}$ at 10~K, respectively. The temperature-dependence of $D$ with the deepest sites blocked is illustrated in \figref{fig:4}. We expect that, under ISM conditions, deep sites will be covered by other species i.e. \ce{H2}, with a consequent increment in $D$.

Diffusion on the surface may be accelerated by quantum mechanical tunneling of the adsorbate. We took that into account in the kMC simulations by correcting the rate constants with the transmission coefficients of Eckart and Bell barriers fitted to the free-energy barriers. The effect by atom tunneling is marginal, even at 10 K: the diffusion constant is increased by a factor of merely 1.3--2.8. Thus, we find tunneling not to be relevant for nitrogen atom diffusion at 10~K, which can be expected from the high mass of N, which agrees with previous results \citep{Pezzella2018} but is at variance with a previous suggestion for oxygen atoms \citep{Minissale2013}. 

Is it justifiable to consider diffusion to be fast under a homogeneous regime, as considered by some theoretical models? A common attempt is to describe diffusion via hopping rates from one minimum to a neighboring one. This may be a justified model if all barriers and all binding sites were similar, like on a crystalline surface. However, even on our comparably small surface model, the barriers of the individual hopping events vary from 0.05 to 7.8~kJ~mol$^{-1}$ with an average of 2.56~kJ~mol$^{-1}$, which result in hopping rate constants with a huge variation from  3.8\e{-30} to 1.1\e{11}~s$^{-1}$ at 10~K. Depending on the surface morphology, some of the higher barriers may be circumvented by the diffusion path. However, the smallest barriers usually merely lead to oscillations between neighboring binding sites rather than to real transport of the adsorbate. Overall, a model that takes the connectivity between binding sites with their realistic barriers into account, like our kMC model, is required to estimate the diffusivity accurately. Effective diffusion barriers ($\Delta F$) should be estimated from such models. 

Most of these arguments have been previously used in the theoretical study of H atoms diffusion on ASW vs. polycrystalline ice \citep{Kuwahata2015, asg17, sen17} and hold here. In the picture presented in \cite{Hama2012}, H atoms block the deep binding sites and recombination of \ce{H2} is possible from another incoming H atom. In the case of nitrogen, however, this mechanism is less likely. Given the similar abundances of H and N in molecular clouds, in the event of complete coverage of binding sites, H atoms diffusion will surpass N atoms diffusion. However, a smaller fraction of N chemistry due to diffusion cannot be discarded under the high-coverage regime.

In the limit of low coverage of ices, we conclude that reactions via the Langmuir--Hinshelwood mechanism with N are extremely unlikely at 10~K.  A closer look at other diffusive mechanisms is warranted in this context. Recently, we have shown that hot-atom diffusion just after exothermic adsorption is a very short-ranged process, with the adsorbate molecules moving only for about 1--2 ps before they become thermalized \citep{Molpeceres2020_2}. They thermalize anywhere on the surface, not necessarily in deep sites. However, deep binding sites are abundant enough to block any significant spatial movement of the adsorbate effectively. As mentioned before, a real surface will be even rougher than our model, so more deep binding sites are expected. After trapping, the adsorbate is effectively removed from the reaction, waiting for an additional reaction partner. 

In the limit of high coverage, on the other hand, diffusion of N could, in principle, proceed. Our diffusion constants at 25~K for high coverage are comparable with those of H at 25~K at low coverage \citep{asg17}, indicating the extreme importance of the number of adsorbates pre-adsorbed on the surface. In experiments, even under the sub-monolayer regime, deep-binding sites are readily occupied. Hence the diffusion of reactive species (such as O or N) is measured as an upper limit of the range of possible diffusion coefficients.

Above which temperature can we expect diffusion to become relevant for thermal diffusion of N on a pristine ASW surface? While it is difficult to assign an absolute number to $D$ sufficient for surface reactivity, we can use the hydrogen atom on ASW as an estimate. $D_\text{H}=1.09\e{-5}$~cm$^2$s$^{-1}$ was obtained at 10~K on ASW \citep{Al-Halabi2007}, a value later claimed \citep{Hama2012} to be somewhat high and recently corrected by the values of \cite{asg17} to $D_\text{H}=5.80\e{-11}$~cm$^2$s$^{-1}$ at 25~K in the limit of low coverage of the surface and  $D_\text{H}=3.30\e{-7}$~cm$^2$s$^{-1}$ at the same temperature when the deepest adsorption sites are occupied. To match the results at 10~K of \cite{Al-Halabi2007} a temperature of about 100~K is necessary, while to match the results of \cite{asg17} at 25 K for a bare surface, a temperature close to 40~K would be necessary.

We finally want to raise awareness of issues arising from the use of average diffusion barriers as fractions of the average binding energy, which is common practice in astrochemical models. From our calculations of the diffusion barrier, we arrived at an average value of 2.56~kJ~mol$^{-1}$, which is 0.76 times the average binding energy. The values that we obtained from the Arrhenius fit in Table \ref{tab:1}, however, show that the effective barrier for diffusion varies between 3.73--6.06~kJ~mol$^{-1}$ (a ratio of 1.1--1.8, using binding energies without zero-point vibrational energies) due to the prevalence of deep binding sites. Such value can decrease below 1.0 for higher surface coverage, as discussed above, meaning that the use of $\Delta F^\mathrm{avg}$ should be only justified in this context. It is worth mentioning, however, that previous modeling studies equivalent to the study presented here found diffusion/binding energy ratios below 1.0 \citep{Karssemeijer2014, asg17}, so further investigation is highly desirable.

\section{Conclusion} \label{sec:conclusion}

In the light of our simulations and in accordance with Pezzella et al. \citep{Pezzella2018}, we conclude that diffusion of nitrogen atoms and implicitly of reactive species heavier and tighter bound by water ice (which, to the best of our knowledge, includes most reactive species) is hindered for a wide range of conditions, and hydrogen diffusion must dominate surface chemistry in these environments. When the surface of the ice is sufficiently populated by other adsorbates, the diffusion of N may be enabled. Alternatively, non-thermal diffusion after hydrogenation can be invoked to explain the formation of molecules such as \ce{CO2} \citep{Ioppolo2011}, organic alcohols \citep{Qasim2019-b}, formaldehyde \citep{Qasim2019} or, very recently, glycine \citep{Ioppolo2020}. We also emphasize here that the importance of non-diffusive mechanisms (Eley--Rideal) should also be re-evaluated in the context of the formation of complex organic molecules (COMs) \citep{Herbst2009, He2017-b}.

\section*{Acknowledgements}

We thank the Deutsche Forschungsgemeinschaft (DFG, German Research Foundation) for supporting this work by funding EXC 2075 - 390740016 under Germany's Excellence Strategy. We acknowledge the support by the Stuttgart Center for Simulation Science (SimTech) and the European Union's Horizon 2020 research and innovation programme (grant agreement No. 646717, TUNNELCHEM). We also like to acknowledge the support by the Institute for Parallel and Distributed Systems (IPVS) of the University of Stuttgart and by the state of Baden-Württemberg through the bwHPC consortium for providing computer time. G. M. thanks the Alexander von Humboldt Foundation for a postdoctoral research fellowship. V. Z. acknowledges the financial support received in the form of a Ph.D. scholarship from the Studienstiftung  des  Deutschen  Volkes (German National Academic Foundation).

\section*{Data Availability}

The data obtained in this article will be shared at reasonable request to the corresponding author. All binding sites and transition states used for kMC simulations can be found in Supplemental Information.




\appendix

\bibliographystyle{mnras}
\bibliography{example} 

\bsp	
\label{lastpage}
\end{document}